\documentclass[aps,showpacs,preprint,superscriptaddress]{revtex4}
\usepackage{graphicx}
\usepackage{subfigure}
\usepackage{float}
\usepackage{amsmath}
\usepackage{amsfonts}
\usepackage{color}
\usepackage{bm}
\usepackage{bbm}
\usepackage{txfonts}
\usepackage{array}
\usepackage{amssymb}
\usepackage{braket}
\begin{document}

\title{Electron-positron pair production in combined Sauter potential wells}
\author{Binbing Wu}
\affiliation{Key Laboratory of Beam Technology of the Ministry of Education, and College of Nuclear Science and Technology, Beijing Normal University, Beijing 100875, China}
\author{Li Wang}
\affiliation{Key Laboratory of Beam Technology of the Ministry of Education, and College of Nuclear Science and Technology, Beijing Normal University, Beijing 100875, China}
\author{B. S. Xie \footnote{Corresponding author: bsxie@bnu.edu.cn}}
\affiliation{Key Laboratory of Beam Technology of the Ministry of Education, and College of Nuclear Science and Technology, Beijing Normal University, Beijing 100875, China}
\affiliation{Beijing Radiation Center, Beijing 100875, China}
\date{\today}
\begin{abstract}
Electron-positron pair production, in combined Sauter potential wells and an oscillating one is imposed on a static Sauter potential, is investigated by using the computational quantum field theory. We find that the gain number (the difference of pair number under combined potentials to the simple addition of pair number for each potential) of the created pairs depends strongly on the depth of static potential and the frequency of oscillating potential. In particular, it is more sensitive to the frequency compared with the depth. For the low-frequency multiphoton regime, the gaining is almost positive and exhibits interesting nonlinear characteristics on both depth and frequency. For the single-photon regime, however, the gaining is almost negative and decreases near linearly with depth while it exhibits an oscillation characteristic with frequency. Furthermore, the optimal frequency and depth of gain number are found and discussed.
\end{abstract}
\maketitle

\section{Introduction}

Quantum electrodynamics vacuum becomes unstable with electron-positron ($e^{+}e^{-}$) pair production in strong background fields \cite{W.Greiner:1985,Sauter:1931zz,Heisenberg:1935qt}. Schwinger \cite{Schwinger:1951nm} obtained the $e^{+}e^{-}$ pair production rate from the vacuum in a strong static constant electric field $E$ by using a proper-time technique, $\exp(-\pi E_c/E)$, where $E_{c}=1.3\times10^{16}\rm V/cm$ is the Schwinger crtital field strength. Since then, the $e^{+}e^{-}$ pairs creation has become a hot research topic \cite{DiPiazza:2011tq,B.S.Xie:2017}.
Two main different mechanisms of created pairs in strong fields are identified, which are close analogies of atomic ionization. They are quantum tunneling mechanism and multiphoton process \cite{E.Brezin:1970,R.Alkofer:2001,Mocken:2010uhp,I.Sitiwaldi:2017}. The Schwinger critical field strength $E_c$, which corresponds to the laser intensity $4.3\times10^{29}\rm W/cm^2$, is so high that the experimental observability of created pairs is very difficult to realize in present laser facilities. However, there are many current construction or planned laser facilities, such as the Extreme Light Infrastructure (ELI) \cite{See:http1}, the Exawatt Center for Extreme Light Studies (XCELS) \cite{See:http2}, the European X-Ray Free-Electron Laser (XFEL) \cite{A.Ringwald:2001}, and the Station of Extreme Light at the Shanghai Coherent Source, can make one expect the experimental observation of pair creation from the vacuum in the future. On the other hand, theoreticians hope to optimize the laser fields to enhance pair production by attempting to use the complex strong fields before the possible experimental observation.

Various theoretical methods have been adopted to deal with this nonperturbative and nonequilibrium problem in pair creation process, such as Wentzel-Kramers-Brillouin (WKB) approximation \cite{Brezin:1970xf,Piazza:2004sv}, worldline instanton technique \cite{Dunne:2005sx,Dunne:2006ur,Schneider:2014mla}, quantum kinetic methods including quantum Vlasov equation \cite{Kluger:1991ib,Abdukerim:2013vsa1,Oluk:2014qta,Sitiwaldi:2018wad} and Wigner function formalism \cite{Hebenstreit:2010vz,Li:2017qwd,Olugh:2018seh}, and so on. Amongst them many works have employed the computational quantum field theory (CQFT) to study not only the pair creation but also some conceptual problems existed in relativistic quantum mechanics \cite{T.Cheng:2010} such as the Zitterbewegung \cite{P.Krekora:2004}, relativistic localization problem \cite{P.Krekora:2004}, Klein paradox \cite{Krekora:2004trv}, and so on.

By the CQFT scheme, some interesting results of pair creation from the vacuum in Sauter potential have been achieved \cite{P.Krekora:2005,Liu:2015mwt,Jiang:2012mwt,Jiang:2013mct,Tang:2013mwt}. For example, the number of pair creation in a strong static well is associated with a population of bound states diving into the negative energy continuum (Dirac sea) \cite{P.Krekora:2005,Liu:2015mwt}. For an oscillating well, the pair creation is caused by the multiphoton process and determined by the frequency of the potential \cite{Jiang:2013mct}. A dynamically assisted Schwinger mechanism, which consists of a strong low-frequency field and a weak high-frequency field in a spatially homogeneous scenario, is found to enhance significantly the rate of created pairs \cite{Schutzhold:2008pz1}. This intrigues many people to consider the different combined fields to enhance the yield of pairs \cite{Li:2014psw, Linder:2015vta, Ababekri:2019dkl}.
The pair production in combining a static Sauter potential barrier and an alternating Sauter potential barrier has been well investigated by the CQFT method \cite{Jiang:2012mwt}. The results show that the pair production in these combined potentials can be increased by several orders of magnitude compared with the production associated with each potential individually. The impact of the static field can accelerate as well as suppress the pair creation process depending on the frequency of an alternating field.
For combined Sauter potential wells, the number of created pairs can also be more than that in a well \cite{Tang:2013mwt}.

Although these previous works about the combined fields have revealed some characteristics for pair production, the impact of depth and frequency of the combined wells on the number of created pairs have not yet been studied completely. In this paper, therefore, we introduce the gain number of the created pairs, which is the difference of pair number under combined wells to the simple addition of that for each single well, to further investigate the effect of combined wells. We focus on effects of the depth of the static well and the frequency of the alternating well on the gaining of pair number. We find that the gain number of created pairs strongly depends on the depth of the static well and the frequency of the oscillating well. For the multiphoton regime, the gain number is mainly positive and exhibits interesting nonlinear characteristics. As the depth increases, it firstly keeps zero, then increases
almost linearly and finally drops with tiny oscillation. For the single-photon regime, however, it even appears negative values for some depths and frequencies. The optimal frequency, mainly lies in the low-frequency multiphoton regime, and depth are found and discussed.

This paper is organized as follows. In Sec. \uppercase\expandafter{\romannumeral2}, we introduce briefly the CQFT framework and our model. In Sec. \uppercase\expandafter{\romannumeral3}, we investigate the gain number in combined wells for different depths of the static well and different frequencies of the
oscillating well by the CQFT and discuss the results. In Sec. \uppercase\expandafter{\romannumeral4}, is a summary of this work.

\section{THE COMPUTATIONAL FRAMEWORK OF CQFT AND THE EXTERNAL POTENTIAL}\label{method}

The number of particles is not conserved in the process of $e^{+}e^{-}$ pair production from the vacuum in strong external fields. It is not accurate to investigate the process with Dirac equation which is a single-particle wave function. In order to better describe the creation and annihilation of electrons or positrons, we employ the CQFT which can provide us with many information about pair production, such as particle number, momentum spectrum, and spatial density distribution at every moment.

In the CQFT, the evolution of field operator satisfies the Heisenberg equation of motion where Hamiltonian is second quantized. Since the number of $e^{+}e^{-}$ pair production is not significant and the force between them is small compared with that of the external electric field, we neglect the fermion interaction. By this assumption, it turns out that the evolution of the field operator $\hat{\psi}(\textbf{r},t)$ also satisfies the Dirac equation in which the vector potential is classical.

In this paper, we use the atomic units (a.u.) as ${\hbar}=m_e=e=1$ and consider an one-dimensional system along the $z$ direction for the sake of simplicity. Here
\begin{equation}\label{Dirac_E}
i\partial{\hat{\psi}(z,t)}/\partial{t}=\left[c\alpha_z \hat{p}_z+\beta c^2+V(z,t)\right]\hat{\psi}(z,t),
\end{equation}
where $V(z,t)$ is the scalar classical external potential along the $z$ direction, $\hat{p}_z$ is the component of the momentum operator along the $z$ axis,  the $\alpha_z$ denotes the $z$ component of the Dirac matrix, $\beta$ denotes unit Dirac matrix, and $c=137.036$ a.u. denotes the speed of light in vacuum. There is no magnetic field in one-dimensional space, so we focus on a single spin. In this case, four-component spinor wave function becomes two components and Dirac matrices $\alpha_z$ and $\beta$ are replaced by the Pauli matrices $\sigma_1$ and $\sigma_3$ respectively.

According to quantum field theory, the field operator can be expanded in term of the time-independent creation and annihilation operators
\begin{equation}\label{Operator}
\hat{\psi}(z,t)=\sum_{p^\prime}\hat{b}_{p^\prime} u_{p^\prime}(z,t)+\sum_{n^\prime}{\hat{d}}^{\dag}_{n^\prime}v_{n^\prime}(z,t).
\end{equation}
The field operator can also be expanded by means of the time-dependent creation and annihilation operators with Bogoliubov transformation
\begin{equation}\label{Operator-T}
\hat{\psi}(z,t)=\sum_{p}\hat{b}_p(t)u_p(z)+\sum_n{\hat{d}}^{\dag}_{n}(t)v_n(z).
\end{equation}
Here $\hat{b}_p$ and $\hat{d}^{\dag}_{n}$ represent the annihilation operator of the electron and creation operator of the positron, $p$ and $n$ are
the momenta of positive and negative energy states respectively, $u_p(z)$ and $v_n(z)$ denote the field-free positive and negative energy eigenstates respectively, and $u_p^\prime(z,t)$ and $v_n^\prime(z,t)$ denote the time evolution of $u_p(z)$ and $v_n(z)$ respectively.
We can express the time-dependent creation and annihilation operators with Eqs. \eqref{Operator} and \eqref{Operator-T} by orthonormality of energy eigenstates of free hamiltonian.
\begin{equation}\label{bpt}
 \hat{b}_p(t)=\sum_{p^\prime}\hat{b}_{p^\prime}U_{pp^\prime}(t)+\sum_{n^\prime}\hat{d}^\dag_{n^\prime}U_{pn^{\prime}}(t),
\end{equation}
\begin{equation}\label{dnt}
\hat{d}^{\dag}_n(t)=\sum_{p^\prime}\hat{b}_{p^\prime}U_{np^\prime}(t)+\sum_{n^\prime}\hat{d}^\dag_{n^\prime}U_{nn^{\prime}}(t).
\end{equation}
where $U_{p,p^\prime}(t)=\bra{u_p(z)}\hat{U}(t)\ket{u_{p^\prime}(z)}$, $U_{p,n^\prime}(t)=\bra{u_p(z)}\hat{U}(t)\ket{v_{n^\prime}(z)}$, $U_{n,n^\prime}(t)=\bra{v_n(z)}\hat{U}(t)\ket{v_{n^\prime}(z)}$, and $U_{n,p^\prime}(t)=\bra{v_n(z)}\hat{U}(t)\ket{u_{p^\prime}(z)}$. The time evolution operator of the field operator $\hat{U}(t)\equiv\
\hat{T}\text{exp}(-i\int_{0}^{t} H d\tau)$, $\hat{T}$ denotes time-order operator, and $H$ is the Hamiltonian of Eq. \eqref{Dirac_E}.
The electronic portion of the field operator is defined as $\hat{\psi}^+_e(z,t)\equiv\sum_p\hat{b}_p(t)u_p(z)$ so that the created electrons' spatial number density can be expressed as
\begin{equation}\label{density}
\rho_e(z,t)=\bra{\text{vac}}{\hat{\psi}^{+\dag}_e}(z,t)\hat{\psi}^+_e(z,t)\ket{\text{vac}}.
\end{equation}
Using Eqs. \eqref{bpt} and \eqref{dnt} and the anticommutator relations $\left\{\hat{b}_p,\hat{b}^\dag_{p^\prime}\right\}=\delta_{p,p^{\prime}}$ and  $\left\{\hat{d}_n,\hat{d}^\dag_{n^\prime}\right\}=\delta_{n,n^{\prime}}$,
the number density of electrons can be rewritten as
\begin{equation}\label{e_density}
\rho_e(z,t)=\sum_n\left|\sum_p U_{p,n}(t)u_p(z)\right|^2,
\end{equation}
where $U_{p,n}(t)$ can be computed by using the split-operator numerical technique \cite{Su:1999mwt}.
By integrating Eq. \eqref{e_density} over space, we can obtain the total number of created electrons as
\begin{equation}\label{N_t}
N(t)=\int \rho_e(z,t)~dz=\sum_p\sum_n|U_{pn}|^2.
\end{equation}

In this paper, using above introduced the CQFT method, we consider a combination of a static and an oscillating Sauter well:
\begin{eqnarray}
V(z,t)=\{V_s+V_o\sin(\omega t)\}S(z),
\end{eqnarray}
where $S(z)=\{\text{tanh}[(z-D/2)/W]+\text{tanh}[(z+D/2)/W]\}/2$, $D$ is the width of well, $W$ is spatial extension of corresponding electric field, and $V_s$ and $V_o$ represent depth of a static and an oscillating well respectively.

For the convenience of study, we use the gain number of created pairs $\Delta N(t)$ to see the effects of the depth and frequency of potential:
\begin{eqnarray}
\Delta N(t) =N_c(t)- N_s(t)-N_o(t),
\end{eqnarray}
where $N_c(t)$, $N_s(t)$, and $N_o(t)$ represent the number of created pairs of combined wells, a static well, and an oscillating well respectively.
In order to shorten simulation time, we consider that electric field turn on and off abruptly, simulated length $L=1.2$ a.u. and simulated time $T=0.002$ a.u.. During this simulated time, the created electrons can not leave the simulation space. Throughout this paper, the characteristic well parameters are chosen as: $V_0=1.47c^2$, $D=10/c$, and $W=0.3/c$.

\section{NUMERICAL RESULTS AND DISCUSSION}\label{result1}

\subsection{Effects of the static potential in the combined wells}

As mentioned in the introduction, the static well can provide some bound states, which
can enhance created pairs. In combined wells, there are
three possible mechanisms to create pairs, which are tunneling, multiphoton and dynamically assisted Schwinger mechanism,
respectively. We investigate the gain number for different depths of the static well. Two fixed frequencies $\omega=1.5c^2$ and $2.5c^2$ are chosen, which correspond to multiphoton and single-photon regime, respectively.

\emph{\subsubsection{Pair production in multiphoton regime ($\omega<2c^2$)}}

\begin{figure}[ht]
\includegraphics[width=0.75\textwidth,height=0.5\textheight]{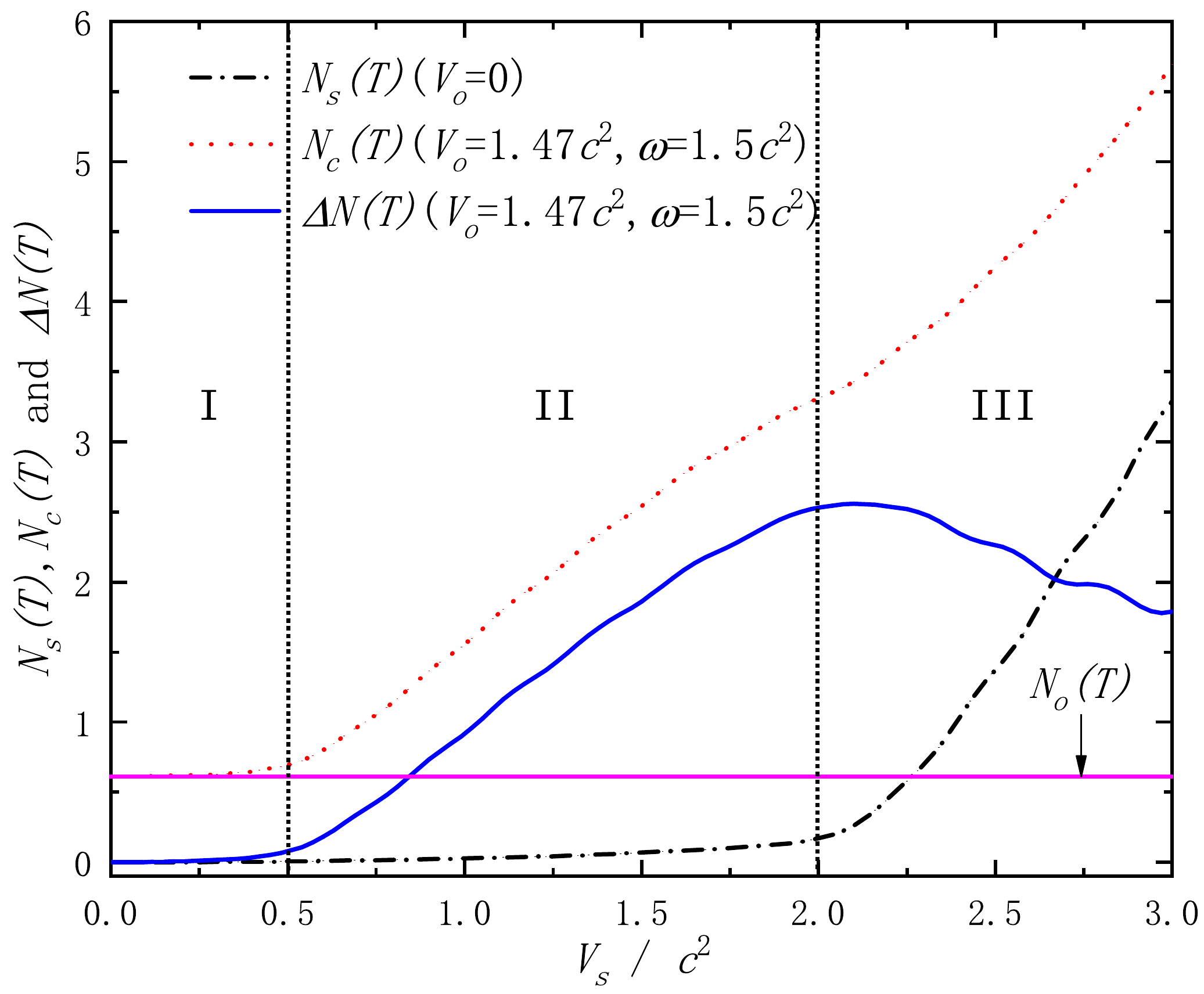}
\vspace{-9mm}
\caption{The final pair number $N_s(T)$ of static well (dot-dashed black), $N_c(T)$ of combined wells (dotted red) and the final gain number~$\Delta N(T)$ (solid blue) at the final time $T=0.002$ a.u. as a function of depth $V_s$. The horizontal solid pink line $N_o(T)$ corresponds to oscillating well. Other potential parameters are $V_o=1.47c^2$, $D=10/c$,$W=0.3/c$, $\omega=1.5c^2$. The simulation time and size are $T=0.002$ a.u. and $L=1.2$ a.u..}
\label{fig:1}
\end{figure}

We first consider the fixed frequency $\omega=1.5c^2$. The final gain number of created pairs $\Delta N(T)$ at final time $T=0.002$ a.u. as a function of depth of the static well $V_s$ between $0$ and $3c^2$ is displayed in Fig. \ref{fig:1}, see solid blue curve. In order to compare with a static or an oscillating well, we also show the other three curves in Fig. \ref{fig:1}.

In Fig. \ref{fig:1}, It can be seen that the number of created pairs $N_o(T)=0.610$ in an oscillating well with the depth $V_o=1.47c^2$ and the frequency $\omega=1.5c^2$ is small, see the horizontal pink line when it is compared with that in the combined wells for most depths of static well. This process can  happen only by absorbing at least two photons and the transition amplitude of more photons would be much smaller obviously due to the multiphoton perturbation characteristic. On the other hand, one can see from the dot-dashed black curve that the final created pair number $N_s(T)$ can be neglected even if its corresponding maximum electric field exceeds Schwinger field $E_c$ ($E_c=c^3$ in atomic units), i.e., $E_{max}=V_s/2W>c^3$ when the depth of static well is less than $2c^2$ ($V_s<2c^2$). It is not surprising because the pair creation needs at least $2c^2$ energy to overcome the gap between positive and negative energy continuum. So created pairs are very small by the tunneling mechanism. Moreover, since the smaller number is triggered by
the process of potential turning on and off \cite{Gerry:2006} so our treatment to turn on and off the potential abruptly in this work is reasonable to some extent.

\begin{figure}[ht]
\includegraphics[scale=0.80]{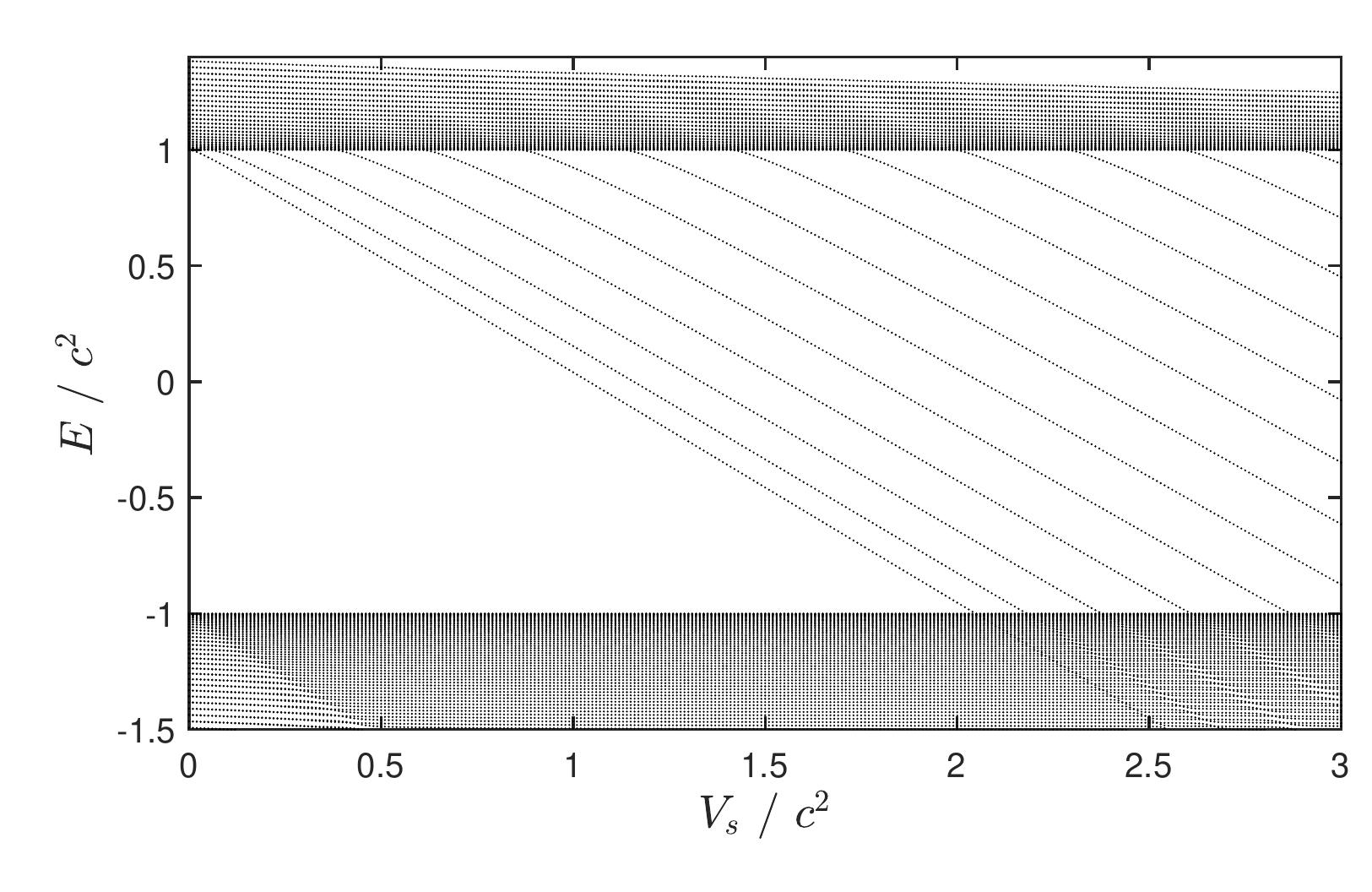}
\vspace{-9mm}
\caption{The energy spectrum of Dirac Hamiltonian in
 external static potential as a function of the depth of the static potential $V_s$. Other potential parameters are given as $D=10/c$, $W=0.3/c$.}
\label{fig:2}
\end{figure}

For $V_s>2c^2$, however, it is interesting that the final created pairs first increase slowly and then almost noticeably linearly improve with the increase of depth. The energy for $V_s>2c^2$ is so sufficient to create pairs by tunneling mechanism. This is also understood by the
bound states diving into the negative energy continuum  \cite{P.Krekora:2005,Liu:2015mwt}. We display the energy spectrum of the total Hamiltonian with external static potential in Fig. \ref{fig:2}.
When the depth of static well is
greater than $2.04c^2$ ($V_s>2c^2$), the bound
states enter the negative energy continuum. These bound states can be realized as a resonance to enhance the created pairs. The deeper the static well, the more bounds states enter into the Dirac sea, which can enhance the number of pair production.
In our case, the final time $T=0.002$ a.u. is so short that electrons can not completely occupy the bound state.
So the final number of created pairs does not tend to the population of bound states diving into the Dirac sea.

Finally, we analyze the two curves in combined wells with an oscillating well $V_o=1.47c^2$ and $\omega=1.5c^2$, see the dashed red curve and the solid  blue curve in Fig. \ref{fig:1}. The curves are divided into three parts: I, II, and III. For part I, the depth of static potential is less than $0.5c^2$. For the depth $V_s<0.33c^2$, the final created pairs $N_c(T)$ stay close to constant and the gain number is about zero. This result shows that the final created pairs $N_c(T)$ are almost irrelevant to the depth of static well. This phenomenon can be intuitively understood from a viewpoint of energy transfer. An electron in the Dirac sea needs at least $2c^2$ energy to
become a real electron. But total energy in combined potentials is $E_t=\hbar \omega +V_s<2c^2$. So the effect of pair production by the dynamically assisted Schwinger mechanism is very weak. Thus the behavior of the combined wells is almost coincident with the result by a single oscillating well. And note that it seems to have the final gain number while it is small within depth $V_s/c^2$ of $(0.33, 0.5)$. This is due to the strong electric field can accelerate created pairs by the multiphoton mechanism to leave the interaction zone to reduce the effect of Pauli blocking. So we can lead to the conclusion that the pairs can be created for the part I is mainly by the multiphoton mechanism.

In part II,  for the depth of static well at about $0.5c^2$ to $2c^2$, however, we can see that both the final total created pairs and the gain number are almost linearly increasing with the depth. The pairs can be created
by the dynamically assisted Schwinger mechanism due to total energy $E_c =\hbar \omega +V_s>2c^2$. This process can be understood that electrons first stay or partly
stay in the bound states, and then they escape from the Dirac sea by absorbing two or more photons \cite{Tang:2013mwt}. We can see from Fig. \ref{fig:2} that
the increasing depth of the static potential provides more bound states for the energy between $-c^2$ and $c^2$, which can be viewed as some ladders to increase the number of created pairs. So both curves in Fig. \ref{fig:1} grow as we expect. For part II, only the number of created pairs by the tunneling mechanism is almost negligible. The number of created pairs by absorbing at least two
photons is constant with increasing the depth of static well. So one can see that the final total pairs and the gain number increase almost with the same slope.
Pairs can be created by the multiphoton absorption and the dynamically assisted Schwinger mechanism. It worth to be noted that the gain number is
greater than the number of created pairs only by multiphoton mechanism for the depth of static well $V_s>0.84c^2$.

Now let us examine the most interesting part III, where the depth $V_s$ lies $2c^2 - 3c^2$. It is very different from the part I and part II that some bound states dive into the negative continuum. Here the created pairs are mainly by the tunneling mechanism. The final created pairs $N_c(T)$ also increase with a nearly constant slope but the slope is a little greater than part II, which is also known from quantum tunneling viewpoint. Next, we show the gain number of created pairs can roughly describe
the number of pairs by the dynamically assisted Schwinger mechanism. Interestingly, the gain number has a peak value $\Delta N_{Max}=2.557c^2$ at about the depth of static well $V_s=2.1c^2$. This result is similar with previous work which consider the relative enhancement $N_c(t)/(N_s(t)+N_o(t))$ in a spatially homogeneous combined electric fields \cite{Orthaber:2011cm}. It is noticed from Fig. \ref{fig:1} that the $N_s(T)$ grows faster than others. The reasonable conjecture is that the bound states diving into the negative energy sea can suppress the process of the dynamically assisted Schwinger mechanism. So we can see that the gain number decreases when the depth is greater than $2.1c^2$.

In summary, for frequency $\omega=1.5c^2$ in the multiphoton regime, the gain number is non-negative and has a peak value in combined wells as the depth of static well increases. Three different parts (I, II and III) for depths of the static well are found and identified for three different dominated mechanisms: the multiphoton process, effective dynamically assisted and the tunneling one.

\emph{\subsubsection{Pair production in single-photon regime ($\omega>2c^2$)}}

\begin{figure}[ht]
\includegraphics[width=0.75\textwidth,height=0.5\textheight]{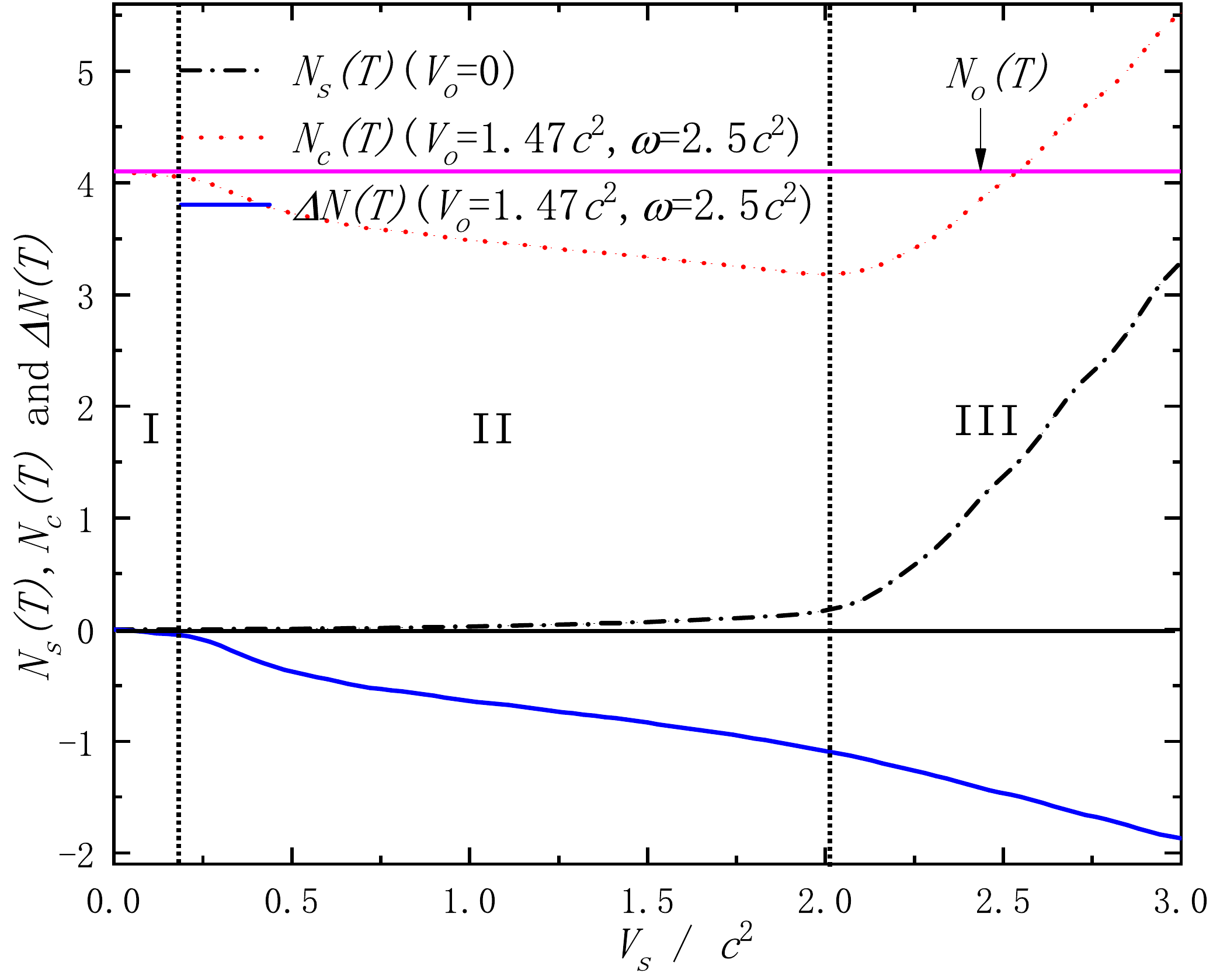}
\vspace{-9mm}
\caption{ The final number of created pairs $N_s(T)$, $N_c(T)$, and $\Delta N(T)$ as a function of depth $V_s$. All parameters are the same as Fig.~\ref{fig:1} except for~$\omega=2.5 c^2$.}
\label{fig:3}
\end{figure}

When $\omega=2.5 c^2$, which produces pairs by absorbing one photon, the numerical results are very different from those of $\omega=1.5 c^2$. The corresponding curves for the pair numbers $N_s(T)$, $N_c(T)$ and $\Delta N(T)$ are plotted in Fig. \ref{fig:3}.

For an oscillating well of the dominating regime for single-photon absorption, the number of created pairs, $N_o(T)=4.099$, increases remarkably compared with the number of the multiphoton sector. On the other hand, the final number of created pairs $N_c(T)$ in combined wells can also be roughly treated as three parts of depth-dependence with the different characteristics, i.e., it keeps a constant value for a low-depth region of $V_s<0.18c^2$ (part I), it is decreasing slowly when $0.18c^2<V_s<2.01c^2$ (part II) and finally, however, it exhibits a rapid increasing when $V_s>2.01c^2$ (part III).

\begin{figure}[ht]
\includegraphics[width=0.75\textwidth,height=0.5\textheight]{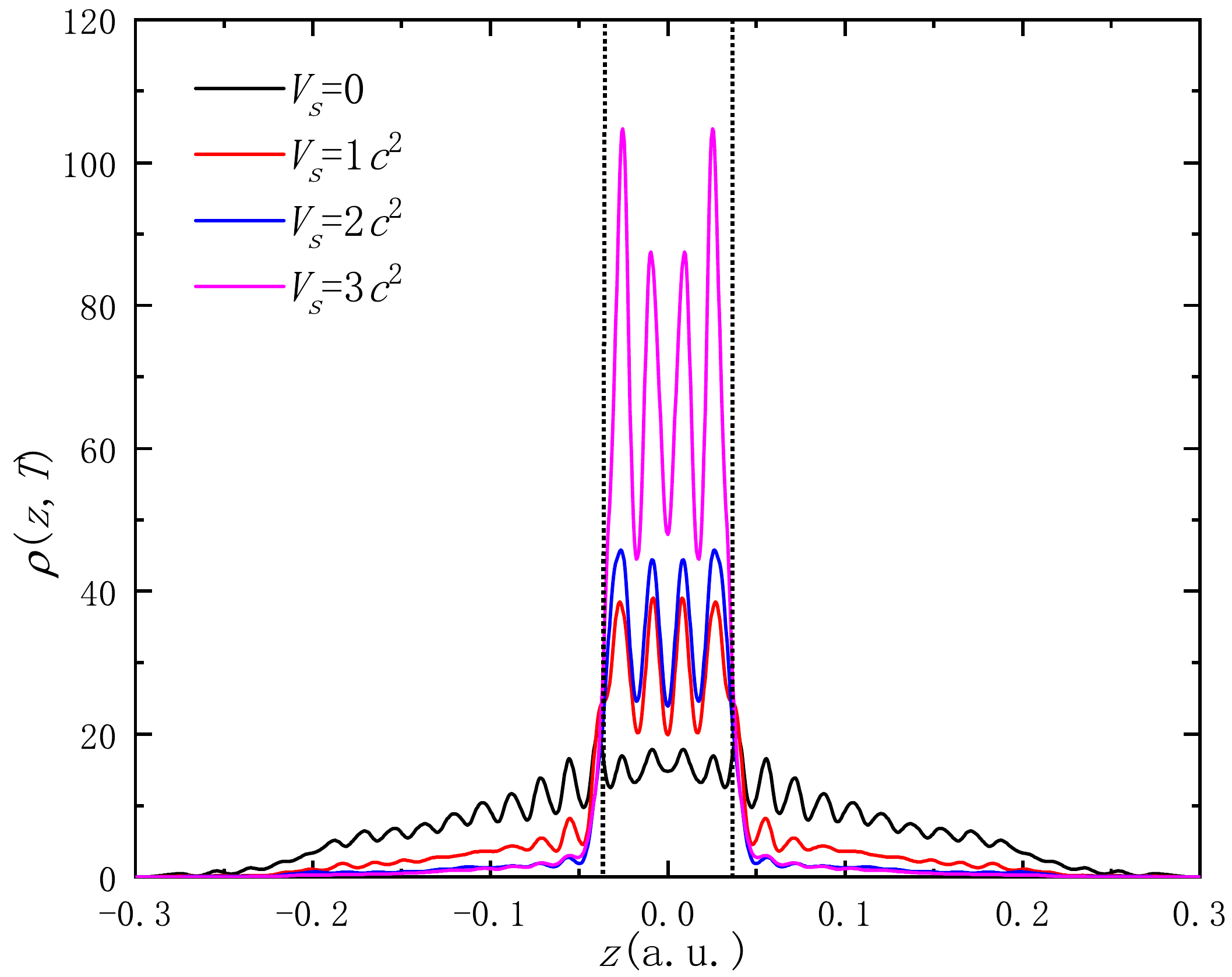}
\vspace{-9mm}
\caption{The final spatial probability density of the created electrons for four different depths $V_s$. Two dotted lines represent the boundary of the well. Other parameters are the same as Fig.~\ref{fig:1} except for $\omega=2.5 c^2$.}
\label{fig:4}
\end{figure}

The range from $0$ to $0.182c^2$ of the red dotted curve in part I is shorter than the corresponding range for $\omega=1.5c^2$ in Fig. \ref{fig:1}.
 The electric field corresponding to the static well is extremely weak so that particles which escaped from the well can not return to the interaction zone to suppress the created pairs. Thus the final number of created pairs $N_c(T)$ keeps a constant value and the gain number is zero as the depth $V_s$ increases.

In part II where the depth $V_s$ lies between $0.182c^2$ and $2.01c^2$, on the one hand, the static electric field is so strong to reverse motion of created pairs moving out of the well, and reduce the number of created pairs due to the Pauli blocking effect. To intuitively understand the effect, we show the final spatial probability density of created electrons for different depths $V_s$ at final time $T=0.002$ a.u., see Fig. \ref{fig:4}. The number of peaks inside the well in the Fig. \ref{fig:4} is the same for different depths, which depends on the frequency. The Pauli blocking effect from the static well is implied in Fig. \ref{fig:4} where the number of created electrons outside the well for the depth $V_s=1.0c^2$ in combined wells is less than the one for an oscillating well. On the other hand, the static well can provide more bound
states between the gap which can enhance the number of pairs. As is expected, the pair number inside the well for $V_s=1.0c^2$ is greater than for an oscillating well. The final number
of created pairs $N_c(T)$ decreases and reaches the minimum value 3.183 at $V_s=2.01c^2$.
The negative gain number suggests that the Pauli blocking effect is more dominant than the effect of bound state in part II.

  Finally, we focus on part III where $V_s$ is greater than $2.01c^2$. One can see from Fig. \ref{fig:4} that the number of created pairs outside the well for $V_s=2c^2$ and $3c^2$ is a nearly constant, which indicates that the reduced number of created pairs due to Pauli blocking effect reaches saturation. The final number of created
pairs inside the well rise remarkably, which is mainly due to bound states diving into the Dirac sea. The final created pairs almost linearly increase in Fig. \ref{fig:3}
which suggests that increased pairs number
from tunneling process is more than the reduced number,
due to the Pauli blocking effect. Moreover, the gain number is also negative and monotonically decreases as the depth of the static well increases.

\begin{figure}[ht]
\includegraphics[width=0.75\textwidth,height=0.5\textheight]{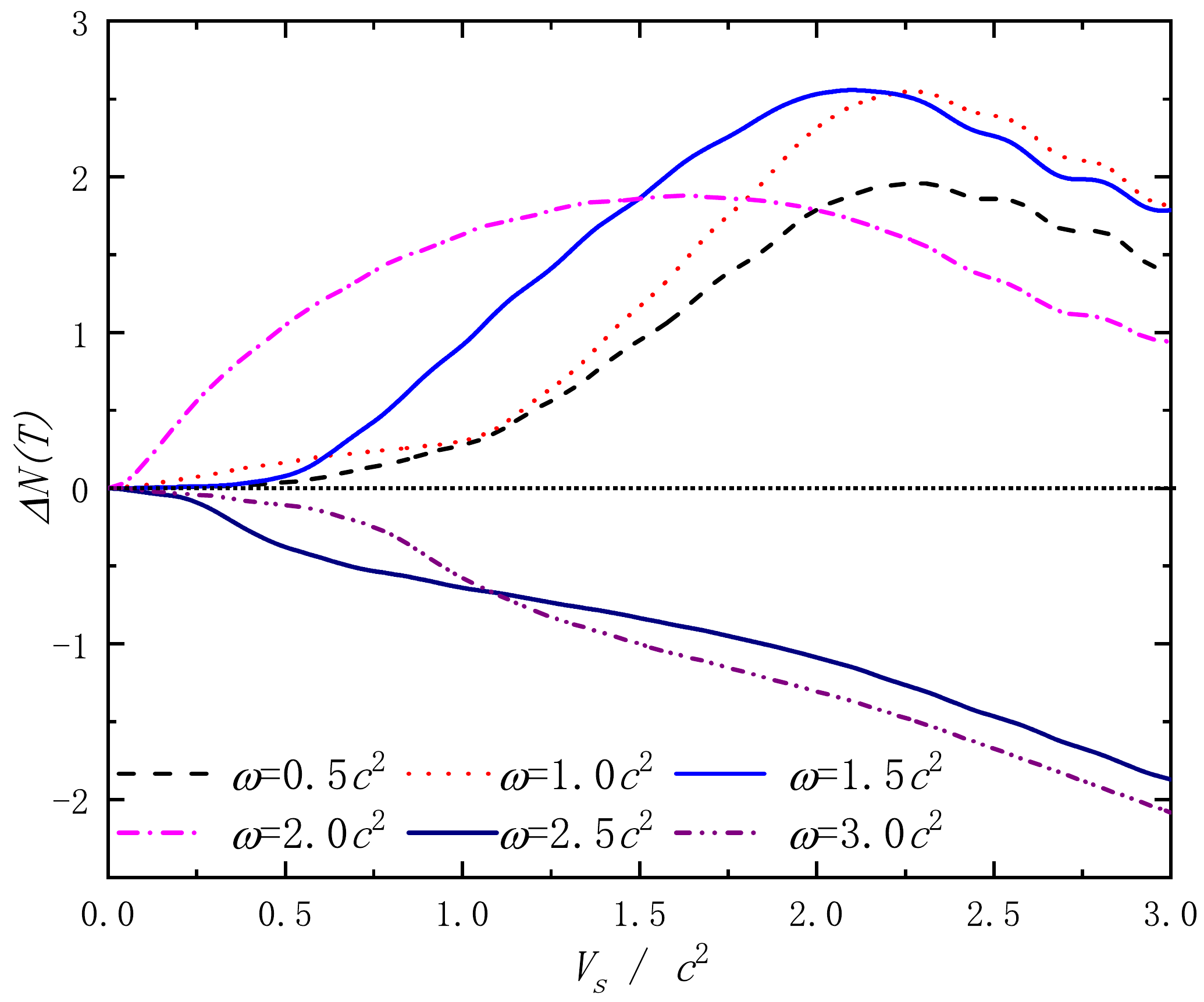}
\vspace{-9mm}
\caption{The gain number of created pairs as a function of the depth of static potential for different frequency.}
\label{fig:8}
\end{figure}

We display the gain number and
its variation with depth for different oscillation frequencies in Fig. \ref{fig:8}. The behavior of the curves is similar to these we discuss above.
It is an interesting phenomenon that as the frequency increases the peak value of the gain number increases first and then decreases.

\subsection{The final pair number as a function of the frequency of oscillating well}

Based on the above analysis, we find that the gain number $\Delta N(T)$ is very sensitive to the frequency. We choose two fixed depths $V_s=0.5c^2$ and $2.5c^2$ to further investigate the effects of frequency.

\emph{\subsubsection{The final pair number in subcritical static potential ($V_s<2c^2$)}}

\begin{figure}[ht]
\includegraphics[width=0.75\textwidth,height=0.5\textheight]{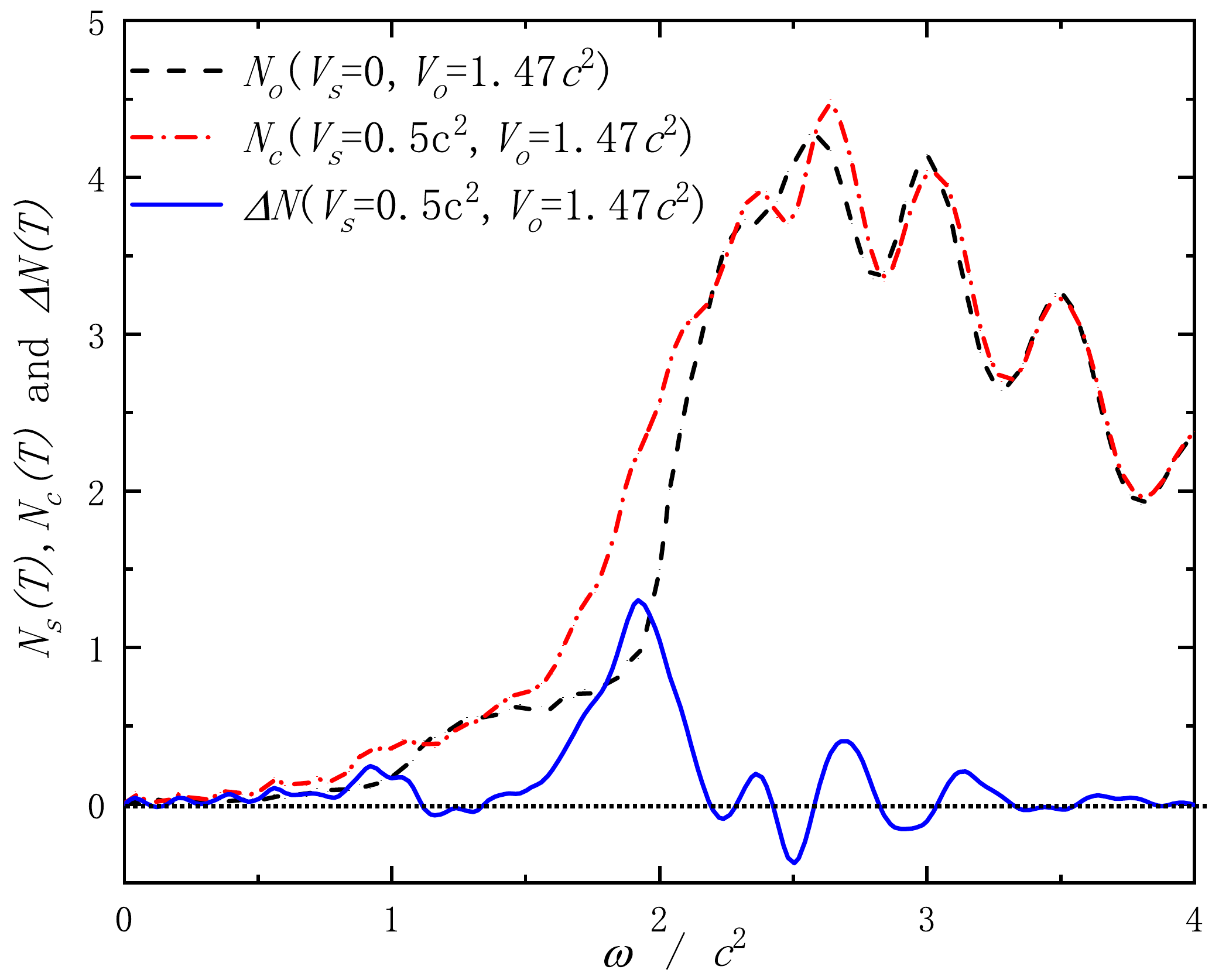}
\vspace{-9mm}
\caption{The final pair number $N_o(T)$ of oscillating well (dashed black), $N_c(T)$ of combined wells (dot-dashed red) and the final gain number $\Delta N(T)$ (solid blue) at the final time $T=0.002$ a.u. as a function of frequency $\omega$. Other potential parameters are same as Fig.~\ref{fig:1} except for  $V_s=0.5c^2$.}
\label{fig:5}
\end{figure}

For $V_s=0.5c^2$, there are no bound states diving into the Dirac sea and the created pairs due to tunneling mechanism are almost negligible.
The corresponding curves for the pair numbers $N_s(T)$, $N_c(T)$ and $\Delta N(T)$ are shown in Fig. \ref{fig:5}.

For single oscillating well, the number of created pairs by multiphoton mechanism increase slowly for $\omega<2c^2$. While $\omega >2c^2$, the number of created pairs increases rapidly due to single-photon absorption and then oscillates damply when $\omega$ is greater
than $2.5c^2$. This oscillation is due to the finite interaction time $T$ and would disappear for $T\rightarrow\infty$ \cite{Jiang:2012mwt}.
The behavior of the pair number $N_o(T)$ is roughly consistent with Ref. \cite{Jiang:2012mwt} and its decrease in the regime of high frequency
can be explained by space-time resolved perspective  \cite{Jiang:2012mwt}.
 The behavior of the pair number of combined fields $N_c(T)$ is almost the same as the pair number $N_o(T)$ with slight differences which is shown by the gain number $\Delta N(T)$ (solid blue curve). The gain number reaches the maximum at $\omega=1.9c^2$ and vanishes at either very low or very high values of the frequency $\omega$ indicating that dynamical assistance becomes more effective around $\omega=2c^2$. Note that the gain number is almost positive when the frequency is less than $2c^2$ with small negative values for the $\omega$ between $1.12c^2$ and $1.32c^2$.

\emph{\subsubsection{The final pair number in supercritical static potential ($V_s>2c^2$)}}
When $V_s=2.5c^2$, the number of pairs from the tunneling mechanism can not be neglected. Also, the corresponding strong electric fields can force the particles to return to the interaction zone, which can reduce the pair number by Pauli blocking effect.

\begin{figure}[ht]
\includegraphics[width=0.75\textwidth,height=0.5\textheight]{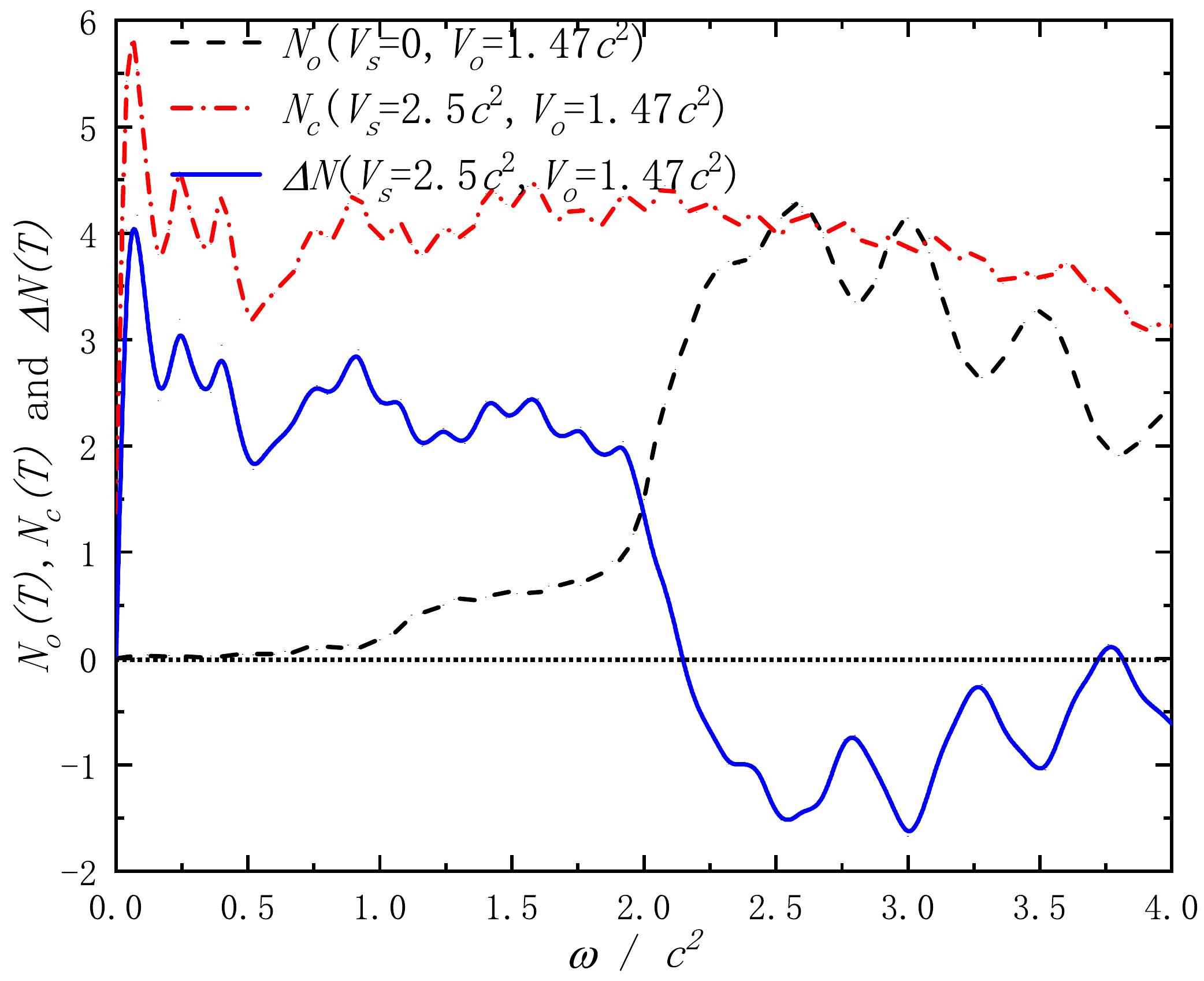}
\vspace{-9mm}
\caption{The final pair number $N_o(T)$, $N_c(T)$ and $\Delta N(T)$ at the final time $T=0.002$ a.u. as a function of frequency $\omega$. All parameters are the same as Fig.~\ref{fig:1} except for the depth of static potential $V_s=2.5c^2$.}
\label{fig:6}
\end{figure}

One can see from Fig. \ref{fig:6} that the number of created pairs $N_c(T)$ and the gain number $\Delta(T)$ are largely different from $V_s=0.5c^2$.
For $\omega<2c^2$, with the increase of frequency, all of them increase rapidly and reach maximum values at $\omega=0.08c^2$ and then slowly fall off with the gradually decreasing amplitude.

In the $\omega<1c^2$ regime, the large gain number $\Delta N$ can be explained by bound states and effective time \cite{Wang:2019oyk}.
During simulated time $T=0.002$ a.u., the depth of combine wells
$V$ grows and reaches maximum value $V_{max}=2.5c^2+1.47c^2=3.97c^2$ at $\omega=0.04c^2$ and then decrease to $1.47c^2$ at $\omega=0.08c^2$ as the frequency increases. The deeper the depth of combine wells, the more bound states diving into Dirac sea generate more pairs. This is the reason why the gain number in the tunneling regime increases rapidly and approaches a maximum value at $\omega=0.08c^2$. When $1c^2<\omega<2c^2$, the large pairs number is interpreted with the more bound states provided by static well between the gap, which can enhance the pair production by dynamical assistance. In the $\omega>2c^2$ regime, the gain number is negative and reaches minimum value $-1.673$ at $\omega=3c^2$, which is due to Pauli blocking effect from the static electric field.

\subsection{The optimal frequency and depth of the static well for the final gain number $\Delta N(T)$}

\begin{figure}[ht]
\includegraphics[width=0.75\textwidth,height=0.4\textheight]{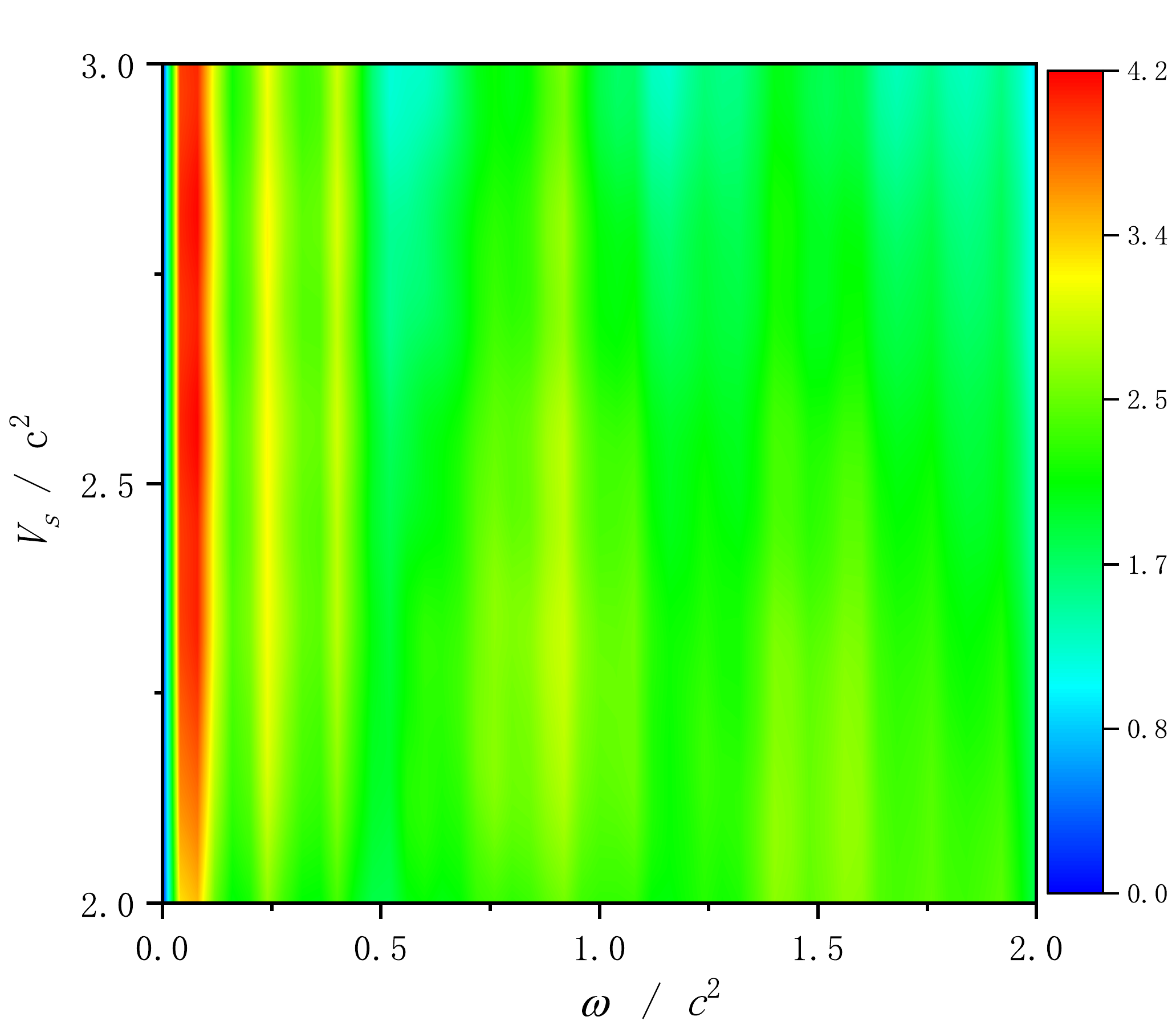}
\vspace{-8mm}
\caption{The contour plot of the final gain number of generated pairs $\Delta N(T)$ at $T=0.002$ a.u. for different parameters ($\omega$, $V_s$), the other  parameters are the same as Fig.~\ref{fig:1}}.
\label{fig:7}
\end{figure}

In this section, we find the optimal frequency and depth of the static well for the gain number. Based on the above analysis, we have plotted the contour plot of the final gain number of pairs for different frequencies ranging from 0 to $2c^2$ and depths of the static
well between $2c^2$ and $3c^2$, see Fig. \ref{fig:7}.

One can see the appearance of bright color bands for some frequencies suggest that the gain number is more dependent on $\omega$ than the static potential depth  $V_s$. The corresponding frequencies of these bands are $0.08c^2$, $0.24c^2$, $0.40c^2$, $0.76c^2$,  $0.92c^2$, $1.06c^2$, $1.40c^2$, and $1.56c^2$ respectively. According to the above frequencies, the optimal depths are provided in Table I.
 The optimal frequency and depth are $0.08c^2$ and $2.58c^2$ respectively, and the tunneling mechanism dominates.

\begin{table}[ht]
\caption{The gain number of optimal depths $V_s$ for different frequencies $\omega$ (in units of $c^2$)}
\centering
\begin{ruledtabular}
\begin{tabular}{ccccccccc}
$\omega$ &0.08&0.24&0.40&0.76&0.92&1.08&1.40&1.56\\
\hline
$V_{s_{opt}}$ &2.58&2.79&2.82&2.34&2.31&2.31&2.07&2.1\\
\hline
$\Delta N(T)$ &4.22 & 3.22 & 3.03& 2.70&3.00&2.57&2.72&2.73\\
\end{tabular}
\end{ruledtabular}
\vskip12pt
\end{table}

\section{Conclusions}

Within the CQFT framework, we have investigated
effects of the depth of static well and the frequency of oscillating well for the gain number of created pairs. And we have obtained the optimal frequency and depth of the static well. The main results can be summarized as follows:

1. In the multiphoton regime, the gain number is almost positive and non-monotonic with increasing depth of the static well.

2. In the single-photon regime, the gain number decreases monotonically and appears negative values as the depth of static well increases.

3. The gain number is more dependent on frequency $\omega$ than the depth of the static potential $V_s$. It reaches maximum value at around $\omega=0.08c^2$, where the optimal depth is $2.58c^2$.

With the increase of the depth of static well,
there are more bound states between positive energy and negative energy continuum, which can enhance the gain number. However, the bound states entering in Dirac sea, Pauli blocking by the static well can reduce the gain number. For the lower-frequency case, the
results can be explained by the bound states diving into the Dirac
sea and effective interaction time.
 For the higher-frequency case, particularly in the single-photon region, the effect of Pauli blocking has a strong inhibitory effect on the gain number. Moreover,
the single-photon process may hinder the channel of
dynamical assistance.
 The assisted mechanism can be further understood by these results. In this work, we only focus on studying the depth of the static well and frequency. To better understand the assisted mechanism, we may also need to consider the width of the potential well.


\begin{acknowledgments}

\noindent
We thank M. Ababekri for his critical reading of the manuscript and helpful remarks.
This work was supported by the National Natural Science Foundation of China (NSFC) under
Grant No.\ 11875007 and No.\ 11935008.
The computation was carried out at the HSCC of the Beijing Normal University.

\end{acknowledgments}


\newpage


\begin{thebibliography}{99}\suppressfloats

\bibitem{W.Greiner:1985}
W.~Greiner,~B.~M$\rm\ddot{u}$ller, and J.~Rafelski,~\textit{Quantum Electrodynamics of Strong Fields}~(Springer Verlag, Berlin, 1985).
\bibitem{Sauter:1931zz}
  F.~Sauter,
  Z.\ Phys.\  {\bf 69}, 742 (1931).

\bibitem{Heisenberg:1935qt}
  W.~Heisenberg and H.~Euler,
  Z.\ Phys.\  {\bf 98}, 714 (1936).

\bibitem{Schwinger:1951nm}
  J.~S.~Schwinger,
  Phys.\ Rev.\  {\bf 82}, 664 (1951).
\bibitem{DiPiazza:2011tq}
  A.~Di~Piazza, C.~Muller, K.~Z.~Hatsagortsyan, and C.~H.~Keitel,
  Rev.\ Mod.\ Phys.\  {\bf 84}, 1177 (2012).
\bibitem{B.S.Xie:2017}
B.~S.~Xie, Z.~L.~Li, and S.~Tang, Matt.\ Radiat.\ Extr. {\bf 2}, 225 (2017).
\bibitem{E.Brezin:1970}
E. Brezin and C. Itzykson, Phys.\ Rev.\ D {\bf 2},1191 (1970).
\bibitem{R.Alkofer:2001}
R.~Alkofer, M.~B.~Hecht, C.~D.~Roberts, S.~M.~Schmidt,  and D.~V.~Vinnik, Phys.\ Rev.\ Lett. {\bf 87}, 193902 (2001).
\bibitem{Mocken:2010uhp}
  G.~R.~Mocken, M.~Ruf, C.~Muller, and C.~H.~Keitel,
  Phys.\ Rev.\ A {\bf 81}, 022122 (2010).
\bibitem{I.Sitiwaldi:2017}
I.~Sitiwaldi and B.~S.~Xie, Phys.\ Lett.\ B {\bf 768}, 174 (2017).
\bibitem{See:http1}
See https://eli-laser. eu.
\bibitem{See:http2}
See http://www. xcels. iapras. ru.
\bibitem{A.Ringwald:2001}
A.~Ringwald, Phys. Lett. B {\bf 510}, 327 (2001).

\bibitem{Brezin:1970xf}
  E.~Brezin and C.~Itzykson,
  Phys.\ Rev.\ D {\bf 2}, 1191 (1970).
\bibitem{Piazza:2004sv}
  A.~Di Piazza,
  Phys.\ Rev.\ D {\bf 70} 053013 (2004).
\bibitem{Dunne:2005sx}
  G.~V.~Dunne and C.~Schubert,
  Phys.\ Rev.\ D {\bf 72} 105004 (2005).
\bibitem{Dunne:2006ur}
  G.~V.~Dunne and Q.~H.~Wang,
  Phys.\ Rev.\ D {\bf 74}, 065015 (2006).
\bibitem{Schneider:2014mla}
  C.~Schneider and R.~Sch$\rm\ddot{u}$tzhold,
  JHEP {\bf 1602}, 164 (2016).

\bibitem{Kluger:1991ib}
  Y.~Kluger, J.~M.~Eisenberg, B.~Svetitsky, F.~Cooper, and E.~Mottola,
  Phys.\ Rev.\ Lett.\  {\bf 67}, 2427 (1991).
  \bibitem{Abdukerim:2013vsa1}
  N.~Abdukerim, Z.~L.~Li, and B.~S.~Xie,
  Phys.\ Lett.\ B {\bf 726}, 820 (2013).

  \bibitem{Oluk:2014qta}

  O.~Oluk, B.~S.~Xie, M.~A.~Bake, and S.~Dulat,
  Front.\ Phys.\  {\bf 9}, 157 (2014).
\bibitem{Sitiwaldi:2018wad}
  I.~Sitiwaldi and B.~S.~Xie,
  Phys.\ Lett.\ B {\bf 777}, 406 (2018).
\bibitem{Hebenstreit:2010vz}
  F.~Hebenstreit, R.~Alkofer, and H.~Gies,
  Phys.\ Rev.\ D {\bf 82}, 105026 (2010).
\bibitem{Li:2017qwd}
  Z.~L.~Li, Y.~J.~Li, and B.~S.~Xie,
  Phys.\ Rev.\ D {\bf 96}, 076010 (2017).
\bibitem{Olugh:2018seh}
  O.~Olugh, Z.~L.~Li, B.~S.~Xie, and R.~Alkofer,
  Phys.\ Rev.\ D {\bf 99}, 036003 (2019).
\bibitem{T.Cheng:2010}
T.~Cheng, Q.~Su, and R.~Grobe, Cont. Phys.{
\bf 51},315 (2010).
\bibitem{P.Krekora:2004}
P.~Krekora, Q.~Su, and R.~Grobe, Phys. Rev. Lett. {\bf 93} 043004 (2004).
\bibitem{Krekora:2004trv}
  P.~Krekora, Q.~Su, and R.~Grobe,
  Phys.\ Rev.\ Lett.\  {\bf 92}, 040406 (2004).
\bibitem{P.Krekora:2005}
P.~Krekora, K.~Cooley, Q.~Su, and R.~Grobe,
  Phys.\ Rev.\ Lett.\  {\bf 95}, 070403 (2005).
\bibitem{Liu:2015mwt}
  Y.~Liu, Q.~Z.~Lv, Y.~T.~Li, R.~Grobe, and Q.~Su,
  Phys.\ Rev.\ A {\bf 91},  052123 (2015).
\bibitem{Jiang:2012mwt}
  M.~Jiang, W.~Su, Z.~Q.~Lv, X. Lu, Y. J. Li, R.~Grobe, and Q.~Su
  Phys.\ Rev.\ A {\bf 85}, 033408 (2012).
  \bibitem{Jiang:2013mct}
  M.~Jiang, Q.~Z.~Lv, Z.~M.~Sheng, R.~Grobe, and Q.~Su,
  Phys.\ Rev.\ A {\bf 87}, 042503 (2013).
\bibitem{Tang:2013mwt}
  S.~Tang, B.~S.~Xie, D.~Lu, H.~Y.~Wang, L.~B.~Fu, and J.~Liu,
  Phys.\ Rev.\ A {\bf 88}, 012106 (2013).


\bibitem{Schutzhold:2008pz1}
  R.~Schutzhold, H.~Gies, and G.~Dunne,
  Phys.\ Rev.\ Lett.\  {\bf 101}, 130404 (2008).
\bibitem{Li:2014psw}
  Z.~L.~Li, D.~Lu, B.~S.~Xie, L.~B.~Fu, J.~Liu, and B.~F.~Shen,
  Phys.\ Rev.\ D {\bf 89}, 093011 (2014).
\bibitem{Linder:2015vta}
  M.~F.~Linder, C.~Schneider, J.~Sicking, N.~Szpak, and R.~Sch$\rm\ddot{u}$tzhold,
  Phys.\ Rev.\ D {\bf 92}, 085009 (2015).
\bibitem{Ababekri:2019dkl}
  M.~Ababekri, B.~S.~Xie, and J.~Zhang,
  Phys.\ Rev.\ D {\bf 100},  016003 (2019).

  \bibitem{Su:1999mwt}
J.~W.~Braun, Q.~Su, and R.~Grobe,
Phys.\ Rev.\ A {\bf 59}, 604 (1999).
\bibitem{Gerry:2006}
C.~C.~Gerry, Q.~Su, and R.~Grobe,
  Phys.\ Rev.\ A {\bf 74},  044103 (2006).
\bibitem{Orthaber:2011cm}
  M.~Orthaber, F.~Hebenstreit, and R.~Alkofer,
  Phys.\ Lett.\ B {\bf 698}, 80 (2011).
\bibitem{Wang:2019oyk}
  L.~Wang, B.~Wu, and B.~S.~Xie,
  Phys.\ Rev.\ A {\bf 100},  022127 (2019).
\end{thebibliography}
\end{document}